# Using NLP to Analyze Political Polarization During National Crises with Twitter

**Parth D. Shisode**
**Portola High School, IUSD**
Irvine, CA 92618
parth.shisode@gmail.com

## Abstract

Democrats and Republicans have seemed to grow apart in the past three decades. Since the United States as we know it today is undeniably bipartisan, this phenomenon would not appear as a surprise to most. However, there are triggers which can cause spikes in disagreements between Democrats and Republicans at a higher rate than how the two parties have been growing apart gradually over time. This study has analyzed the idea that national events which generally are detrimental to all individuals can be one of those triggers. By testing polarization before and after three events (Hurricane Sandy [2012], N. Korea Missile Test Surge [2019], COVID-19 [2020]) using Twitter data, we show that a measurable spike in polarization occurs between the Democrat and Republican party. In order to measure polarization, sentiments of Twitter users aligned to the Democrat and Republican parties are compared on identical entities (events, people, locations, etc.). Using hundreds of thousands of data samples, a 2.8% increase in polarization was measured during times of crisis compared to times where no crises were occurring. Regardless of the reasoning that the gap between political parties can increase so much during times of suffering and stress, it is definitely alarming to see that among other aspects of life, the partisan gap worsens during detrimental national events.

## 1    Introduction

Currently, the COVID-19 pandemic is posing a danger to the welfare of virtually every single nation. However, as noted in Bol et al. 2020 [1], one thing generally measured to be improved is a population's relationship with its government, given that it is a democracy. In the context of COVID-19, they also found a general trend of increased approval for the current government-affiliated political party, as well as government authorities in general. It should be stated that on average, the majority of a population also approves of a government's response to the COVID-19 epidemic.

However according to the Pew Research Center [2], this has not been the case for the United States, where 65% of Americans President Trump was too slow in the initial national COVID-19 response. Attitudes regarding restriction different significantly between political parties: 81% of Dem./Dem.-Leaning Americans are concerned about public restrictions being lifted too quickly, while the same figure is 51% for Rep./Rep.-Leaning Americans.

The purpose of this research study is to explore the degree to which levels of bipartisanship in the US have changed during times of crisis, beyond background change which has already been occurring. Our hypothesis is that during times of crisis, levels of partisanship in the United States will increase, with both Democrats and Republicans becoming more polarized on average in the messages they send via Twitter. Partisanship is defined in this study using sentiment differences around a singular, concrete set of entities; if the perception of a single set of facts still leads to different interpretations and opinions, then bias and polarization are responsible for those differing sentiments. As to how polarization and partisanship will be measured, Twitter has proved to be a resourceful tool in scraping sentiment of the public on entities such as popular organizations, trending celebrities or political figures, geographic regions, and even Twitter handles (Ringsquandl et al., 2016 [3]). The crises tested will be non-acute, in that they did not occur during one singular day or moment: 2012 Hurricane Sandy, 2019 North Korean Missile Test, and lastly, 2020 COVID-19 pandemic. Non-acute events were chosen because it would be more likely to see bias develop over time, once information regarding the event is widespread over a considerable time length. Additionally, these events were chosen due to the fact that the Twitter data surrounding these time periods was accessible relative to other events. Prior research has shown that there is often a uniform sense of grief after trauma and tragedy, typically resulting in solidarity throughout nations (Kropf et al. 2014 [4]). It is interesting that we didn't see this reaction from the American public during the U.S. government's response to the COVID-19 pandemic, given the difference between how Democrats and Republicans still disagree on government-mandated restriction due to COVID-19 [1].

    We classify political perspectives of a Twitter user by comparing the number of Democrat political figureheads with the number of Republican political figureheads that they follow. From Demszky et al. 2019 [5], we have received a "follower list" of Democrat and Republican figureheads, used to classify the political perspective of a tweet based on its user. Using CoreNLP, entities were then extracted from tweets into a dataset alongside entity metadata including corresponding sentiments from the users. Then, polarity for every individual entity with sentiments from both Democrats and Republicans is calculated by analyzing the difference in sentiment between the two political parties. The polarity across all entities as a single value is then calculated by taking a weighted average of the polarities for every individual entity. As a result, it is possible to examine how a single set of events and facts can be seen through a different lens based on political party. This process is further touched upon in the "Methodology" section of this study.[1]

    After analysis of the results, it was determined that polarization during a national detrimental event can be greater than the polarization between political parties that has already been expected to occur. It was also very interesting to note that for the majority of measured sentiments, Republicans tended to have a lower sentiment on average than Democrats. The "baseline" tweets were taken for a period of 3 weeks before the event started, while the "crisis" tweets were taken for a period of 3 weeks after the event started; the goal here was to provide an equal timeframe to account for the fact events may take time in order to for the effect of changing sentiment to take place.

## 2    Related Work

The hypothesis of the "partisanship perceptual screen" plays a critical role in how this study chooses to structure its methodology, as well as its overall goal. According to Mary McGrath, a perceptual screen is capable of "causing adherents of opposing parties to perceive different information from the same set of facts" (McGrath 2016 [6]). This has been based on the research of Gerber and Huber, analyzing the effect of the partisanship perceptual screen on real-world economic behavior (Gerber et al., 2009 [7]). These papers served to aid in understanding how to analyze the set of tweets and define partisanship; in this study, the same specific entities only with measurable sentiment from both the Democrat and Republican are considered.

    This study is also examining the effect of "unity after tragedy" a social phenomenon where during/after a tragic event, the victims and those surrounding victims are able to display social solidarity (Hawdon, J., & Ryan, J. 2008 [8], Sweet 1998 [9]). The hypothesis surrounding our study states that this phenomenon may not occur in the United States between members of the Democrat and Republican Party. Therefore, this phenomenon is undergoing validation through this study. According to the Pew Research Center [10], a further gap of political ideology occurred between Democrats and Republicans once Donald Trump initiated his response to the COVID-19 pandemic. This raises the question: does the "unity after tragedy" phenomenon still occur when

---

[1] The program code and information necessary for this project can be found at the following web address: *https://github.com/21ShisodeParth/partisanship*

the differences between individuals are in political values?

A research study with a similar purpose was that of Demszky et al. [4], analyzing political polarization regarding mass shootings, with data being fed from Twitter. While this study does take a more complex approach to understand the intricacies of how Democrats and Republicans react differently to a significant tragic event, the goals of this study are aligned with that of this paper. While this paper chooses to examine topic choice, framing, affect, and illocutionary force, this study focuses almost solely on identifying whether a partisan perceptual screen continues to exist and affect how Americans view the same set of facts surrounding tragic incidents. Additionally, the way the Demszky et al., chooses to calculate partisanship (derived from the leave-out estimator in Gentzkow et al. forthcoming [11]), has a large influence on the calculation method in this study. Partisanship is measured here based on existing values of polarization calculated for sets of entities which contains measurable sentiment from both political parties. Additionally, partisan assignment is performed using the same method. A list of Twitter handles of Democrat and Republican figureheads was supplied by Demszky et al., along with a list of IDs of every single follower the political figureheads have. Analyzing whether an individual follows more Democrat or Republican political figureheads led to a corresponding party assignment.

# 3      Methodology

The hypothesis of this study is that partisanship will increase in times of crises. A sub hypothesis is that the difference in partisanship of different parties will be even greater than background differences due to an ever-growing gap in political values between Democrats and Republicans, according to a study by the Pew Research Center.

## 3.1      Core Definitions

The definition of partisanship in this study is based on the partisan "perceptual screen", the idea that surrounding a single entity or set of facts are multiple sets of opinions, as a result of different interpretations based on different political parties (McGrath 2016 [6]). As a result, the definition of partisanship in this study will be a measurement of disagreement based around the differing sentiment of concrete entities.

The definition of crisis in this study will be based on a definition from IGI Global: "A situation or time at which a nation faces intense difficulty, uncertainty, danger or serious threat to people and national systems and organizations and a need for non-routine rules and procedures emerge accompanied with urgency". The crises in this study as listed will be the 2012 Hurricane Sandy, 2019 North Korean Missile Test Surge, and the 2020 COVID-19 pandemic. These crises were chosen because they were not "acute tragedies", but rather long-term events; the response to Hurricane Sandy lasted multiple months, North Korea had several tests over the span of many days, and COVID-19 has been infecting individuals from March 2020 to present, over many months.

## 3.2      Assumptions

For this research study, it is necessary for every tweet to hold a political party assignment of Democrat, Republican, or neither. This is crucial in understanding the perspective and background that comes alongside every sentiment of a tweet. An assumption being made is that a person has a binary party affiliation, not taking into consideration that a person could identify as a moderate or party-leaning individual. In order to assign a party, we simply reference the accounts they are following and based on whether a user follows more Democrat or Republican politicians or figureheads, we are able to assign a class. This is a technique alike that of Volkova et al. 2014 [12], and Demszky et al. (2019), with the latter stating that this method or party assignment "takes advantage of homophily in the following behavior of users on Twitter" (Halberstam and Knight, 2016 [13]).

Additionally, it is assumed that external national political events during the period being studied would not skew partisanship significantly. Examples of these would be state elections or national elections as well as any national presidential debates. The timeframes here are only in respect to the events tested (Hurricane Sandy, N.K. Missile Testing, COVID-19) and do not take any other events in the same time period into account.

Lastly, it is assumed that in the statistical definition, the automated sentiment classifier being used (Stanford CoreNLP) is not biased, and will be able to yield accurate results. This sentiment classifier has been previously trained on movie reviews, able to classify between "very negative", "negative", "neutral", "positive", and "very positive." There is an assumption that this model will be unbiased in this study, where it is required to predict sentiment on text in a political context. It is also assumed that the sentiment of an entire sentence applies to every sentiment located in the sentence; this is how sentiments are assigned to entities.

### 3.3   Data Pipeline

From Dora et al. 2019, we've received a list of Twitter handles of Republican and Democrat political figureheads. Additionally, Dora et al. has supplied us with a list of Twitter User IDs that follow every single one of these political figureheads. Every single event chosen has two timeframes we've selected— a baseline period and a crisis period. The baseline period is 3 weeks prior to the point where the event gains large amounts of traction, both in terms of the public (via social media, online posts, etc.) and the government's acknowledgement of the respective event. For every single event, the specific chosen time frames are displayed in the table below. In order to clarify the role of how entities are utilizes in the data pipeline and computation, information about entity types, instances, and mentions need to be elaborated upon. The following entity types were accepted by the CoreNLP annotator to be classified as tweets: ['LOCATION', 'MISC', 'PERSON']. An example of an entity instance would be "Donald Trump", or additional instances of the previously mentioned entity types.

The overarching goal of this study is to assess the polarization surrounding entities from the Democrat and Republican political parties; single entities with opposing sentiment between the two political parties represent partisanship. The input data itself is structured as a tweet object, a dictionary-style data structure which contains information about the user's account, text of the tweet, and time posted surrounding the tweet. In order to transform these objects into a data form that can be analyzed to create a single polarization value, the tweet objects will need to be converted into CSV files which contain a collection of entities alongside its associated Democrat and Republican sentiments. In order to perform this data transformation, 4 steps are taken. Steps 2 - 4 are completed separately for all tweets with a "Democrat" standpoint and all tweets with a "Republican" standpoint.

### 3.3.1   Tweet Annotation with CoreNLP

CoreNLP is a linguistic annotator able to provide information on a text regarding sentiment, named entities, dependency, and parts of speech, as well as other aspects not utilized within this study. The input required for this step is the tweet's text, while the result is a dictionary style object with information on the aspects of the text we'd like to examine or could possibly reference in the future (sentiment, NER, dependency, POS for this study) for every entity instance. It should be noted that past deleted tweets were included in the dataset alongside current tweets; deleted tweets were not annotated and used in the study at all.

### 3.3.2   Partisan Assignment to Tweet

From the information regarding the names of figurehead Twitter handles and respective follower IDs provided by Dora et al., we are able to assign a political party to each tweet based on the user that posted it. For each ID, we search the number of times it appears in a list of followers for each Democrat figurehead, labeled value $f_d$. This process is replicated for each Republican figurehead, yielding $f_d$. If $f_d = f_r$ (including $f_d = f_r = 0$), then the tweet will not be used, as only tweets with a partisan bias are useful in this study.

### 3.3.3 Data Creation for Entity Referenced in Tweet

After each named entity *e* is identified inside the tweet's text, a new datum is created. It should be noted that in order to use to most relevant input data, entities tagged by CoreNLP as the following types were discarded and not used to create data: ("EMAIL", "DATE", "NUMBER", "PERCENT", "TIME", "MONEY", "URL"). There is an assumption that entity types that are events, locations, organizations, and people are more likely to create any meaningful measurement of polarization. For the list of tags of entities not used, it does not seem plausible that these types of entities create any meaningful difference of opinion between Democrats and Republicans. A single datum is then created which contains the following elements: the name of the entity, the user ID of the entity's original tweet, the entity sentiment (calculated using the general sentiment of the entity's original sentence), and the associated political party of the tweet. This is repeated for every entity derived from the source of tweets. Each datum is then compiled as a single row of one CSV file, with the primary key being the entity name.

### 3.3.4 Transformation of Entity Data CSV into Dictionary

Entities that appear in multiple tweets will inevitably appear in multiple rows of the CSV containing the data of the entities compiled from the input of tweets. Having entities in multiple rows does not allow for calculation of average sentiment, which is why this CSV has to be transformed into a new dictionary format. The keys are the of name of each entity $e_t$, while the values are a list of two elements (average sentiment *s*, number of mentions from original CSV *n*). In order to ensure that entity names that differ only because of capitalization (ex. "Donald Trump" vs "donald trump") are considered, all entity names are made lowercase. If two entities from the CSV are found to be identical, the dictionary value for this entity is recalculated to take the average sentiment *s* of both entities, and the number of mentions *n* is updated to represent the number of repeats. An entity mention would be a version of an entity instance that has been repeated; for example, "Donald Trump" was to be found in several different tweets. Essentially, the multiple entity mentions are being reduced to entity instances as a result of this step.

### 3.4 Computation

### 3.4.1 Polarization Across a Single Entity

Once the two dictionaries, one for each political party, are compiled containing the average sentiments on entities and the number of times it appeared throughout the data input, we can then calculate polarization using the formula below. This calculates the amount of polarization *p* across a single entity.

$$p = \frac{|s_d - s_r|}{5} \quad (1)$$

While $s_d$ represents the average sentiment across an entity from users identified as Democrats, $s_r$ represents the counterpart for users identified as Republicans. In order to normalize the value of polarization to be between 0 and 1, the difference of $s_d$ and $s_r$ is divided by 5. The scale for sentiment analysis used by CoreNLP ranges from 0 (very negative) to 4 (very positive), allowing 5 possible options for sentiment value.

### 3.4.2 Polarization Across Multiple Entities

In order to calculate polarization across all entities for the time frame of an event, we can utilize $p_i$. The value $p_{total}$ will be utilized later to measure the presence of partisanship. It is calculated using a weighted average of the $p$ values and the number of mentions the entity itself has had in total, summing $n_d$ and $n_r$.

$$p_{total} = \frac{|\sum_0^i [p_i \times (n_d + n_r)]|}{(n_d + n_r)} \quad (2)$$

| *Event Response* | Volume of Tweets Tested - Baseline | Volume of Tweets Tested - Crisis | Time Period – Baseline | Time Period – Crisis | Total Number of Entities - Baseline | Total Number of Entities - Crisis |
|---|---|---|---|---|---|---|
| Hurricane Sandy - 2012 | 203,409 | 203,462 | 9/24/12 - 10/15/12 | 10/22/12 - 11/12/12 | 4920 | 4963 |
| N. Korea Missile Test Surge - 2019 | 210,612 | 209,981 | 10/3/19 - 10/24/19 | 10/31/19 - 11/21/19 | 4795 | 4680 |
| COVID-19 Pandemic - 2020 | 264,054 | 263,503 | 1/31/20 - 2/21/20 | 2/28/20 - 3/20/20 | 5216 | 5191 |

Table 1: Volume of tweets and entities, time period of baseline and crisis per national crisis

## 4 Results

| *Event Response* | Avg. Dem. Baseline Sentiment | Avg. Dem. Crisis Sentiment | Avg. Rep. Baseline Sentiment | Avg. Rep. Crisis Sentiment | Polarization - Baseline | Polarization - Crisis |
|---|---|---|---|---|---|---|
| Hurricane Sandy - 2012 | 1.87 | 1.82 | 1.84 | 1.82 | 1.8% | 2.5% |
| N. Korea Missile Test Surge - 2019 | 1.72 | 1.70 | 1.71 | 1.67 | 2.0% | 4.9% |
| COVID-19 Pandemic 2020 | 1.86 | 1.79 | 1.84 | 1.77 | 2.1% | 6.9% |

Table 2: Democrat/Republican baseline and crisis sentiment and polarization per crisis

It can be observed from the table that in all three events which were observed, the polarization from the crisis time period to the baseline time period increased. This seems to be strong evidence that in times of crisis, the average polarization between the two parties increases. On average between the three events, the polarization between baseline and crisis increased by 2.8%, from 2.0% to 4.8%. It is possible that the slight increase in baseline polarization from 2012 to 2019 represents the ever-growing gap between the Democrat and Republican parties. Additionally, it is interesting to note that on average, the Republicans held a slightly more negative sentiment than Democrats when referring to entities for baseline and crisis periods of all events. The single exception to this is the average Democrat and Republican sentiments during hurricane Sandy. An unexpected outcome from this data was the slightly lower sentiment in regard to the N. Korea Missile Test Surge. In terms of pure consequence and resources lost as a result of a detrimental event, the North Korea Missile Test Surge is significantly lower than that or Hurricane Sandy and COVID-19. In the U.S. Hurricane Sandy caused destruction of infrastructure, a $70 billion dollar loss, and 125 deaths. The COVID-19 pandemic has caused 213,000 deaths at the time of the writing of this paper, and multiple trillions of dollars of loss for the U.S. government and citizens. Meanwhile, the N. Korea Missile Test Surge hasn't had nearly as significant an effect on the Americans population. Yet, the overall sentiment is lowest for this time period.

## 4.1 Discussion

During the course of this study, there were limitations which may have an effect on the data collection, leading to potential slight inaccuracies in the polarity value calculations.

When entities were being collected, alongside their respective sentiments, they were simply selected if the CoreNLP tool was able to classify it as a location, organization, or person. However, there was no implementation of a method to ensure that each entity was relevant to the political world or political ideologies. To account for this, there is an assumption that all entities mentioned in a tweet are being chosen based on the viewpoint of a political party. For example, the entity keywords "Trump" and "Bernie" are politically relevant terms, and are likely to garner opposing sentiment. However, the entity "Arnold Schwarzenegger" may lead to differences in sentiment that may not be directly apparent. The assumption, for example, would be that Arnold Schwarzenegger is a Democrat, or that another confounding variable leads him to receive a higher sentiment from users labeled as a Republican.

Due to the fact that in terms of percentage-points, the polarity between Democrats and Republicans here appears to be less than that listed by the Pew Research Center, it may be possible that individuals are less likely to express a political bias on social media. Compared to the data collection method Pew Research Center employs (individually-completed surveys), there simply is not the same amount of opportunity to express one's political opinion through bias. For example, adults are specifically asked about issues such as gun control legislation, while directly indicating their political preference. This would directly lead to this data specifically existing for the purpose of understanding differences in political preference. Twitter data, however, does not fulfill this purpose due to the fact that it is a social media platform.

## 4.2 Future Work

In the future, it would be beneficial to examine the differences in sentiment within the same party. In this study, two data sets of entity sentiments were used, with one set coming from Democrat users and the other from Republican users. However, it would be very valuable to assess the validity of this study by replicating it using two data sets from the same political party. If the difference in polarities from the study compared to the altered study is statistically significant, then this study's methodology will need to be revised in order to account for this. Additionally, as mentioned as a limitation, it would be useful to include an algorithm that is capable of detecting the amount of relevance of an entity to the topic which is being assessed. For example, in this study the topic would be the Democrat and Republican parties, as well as any entities that have significance in the political world.

Understanding how the sentiment surrounding a single entity changes over time is also valuable. Analyzing how the amount of time that an event has an effect on the population's general change in sentiment is very valuable, and a good future research question. Filtering the popularity of entity changes more heavily can lead to ultimately more accurate predictions; for the sake of low data availability during this study, this aspect was not emphasized. However, with the presence of more data, only testing data with a high number of mentions from both parties can be useful. When taking a look at the entities, it will be useful to analyze what the study would look like if only a 'PERSON" entity was recognized, and whether it truly is different from the data obtained during this study.

Although this study was conducted with the Democrat and Republican parties based in the United States, these results should be applicable to other nations that follow a multi-party political culture. Applying this study and analyzing how a foreign population's polarization can change after an event in another country may further validate the results of this study and show that the results can be applied on a broader scale. In addition, citizens of other countries may not necessarily use Twitter and may use other social media. Even in the United States, it would be very useful to analyze data which may come from different sources such as a political forum, which may even aid in making the data more politically relevant. Changing the country or the social media source of this study's data may change the way polarization occurs, because of a potential change in user demographics such a region and average age. Even in the United States, for example, understanding user demographics would be important. On average, more rural and older citizens tend to align with the Republican party, while younger voters in urban areas tend to align with the Democrat party.

# 5 Conclusion

After collecting data both before and after the start of a national detrimental event, it's been deduced that such an event can increase polarization between political parties, such as the Democrat and Republican parties, beyond any background polarization which has already been occurring. By the scale used in this study, the increase in polarization increased at a rate of 2.8% on average between political parties during time periods of national crisis. By utilizing Twitter data of users that identify as Democrats and Republicans and analyzing their sentiment on common entities, polarization can be analyzed. If the sentiment towards common entities differs significantly between Democrats and Republicans, a high polarization exists. Political parties for individuals are deduced in this study based on whether they are following more Democrat or Republican politicians/figureheads on Twitter. Polarization data was collected on Democrat and Republican users for the following three events (Hurricane Sandy [2012], N. Korea Missile Test Surge [2019], COVID-19 [2020]). Sentiment/polarization data was calculated using the Stanford CoreNLP tool, accessed through Python. Due to the fact that this data was in fact taken through Twitter, political relevance of the entities tested could possibly be improved in future studies, where an alternative could be through a politically-based social media or forum. Significant future work could also include a replication of this on multi-party systems outside of the United States. If the results still hold true, then this study's results would be validated for applicability outside the United States.

## Acknowledgments

This paper was written under the mentorship of Gabor Angeli. I would like to thank him for his helpful guidance during the research process. Additionally, I would like to thank Dora Demszky for aid in handling Twitter data for this project.